\magnification=\magstep1
\overfullrule=0pt
\parskip=6pt
\baselineskip=15pt

\def\PD{{\cal P}}
\def\La{{\Lambda}}
\def\la{{\lambda}}
\def\al{{\alpha}}
\def\si{{\sigma}}
\def\dl{{\delta}}
\def\ep{{\epsilon}}
\def\om{{\omega}}
\def\af{{A_r^{(1)}}}
\def\tr{{\tilde\rho}}
\def\tk{{\tilde k}}
\def\ct{{\it T}}

\def\arx{\hbox to 8mm{\rightarrowfill}}
\def\ary{\hbox to 10mm{\rightarrowfill}}


\input epsf
\headline={\ifnum\pageno>1 \hss \number\pageno\ \hss \else\hfill \fi}
\pageno=1
\nopagenumbers
\hbadness=100000
\vbadness=100000

\centerline{\bf PERMUTATION WEIGHTS FOR $\af$ LIE ALGEBRAS }
\medskip

\centerline{\bf Meltem Gungormez} \centerline{Dept. Physics, Fac.
Science, Istanbul Tech. Univ.} \centerline{34469, Maslak, Istanbul,
Turkey } \centerline{e-mail: gungorm@itu.edu.tr}

\medskip

\centerline{\bf Hasan R. Karadayi} \centerline{Dept. Physics, Fac.
Science, Istanbul Tech. Univ.} \centerline{34469, Maslak, Istanbul,
Turkey } \centerline{e-mail: karadayi@itu.edu.tr} \centerline{and}
\centerline{Dept. Physics, Fac. Science, Istanbul Kultur University}
\centerline{34156, Bakirkoy, Istanbul, Turkey }

\medskip
\centerline{\bf{Abstract}}
\vskip 5mm

When it is based on Kac-Peterson form of Affine Weyl Groups,
Weyl-Kac character formula could be formulated in terms of Theta
functions and a sum over finite Weyl groups. We, instead, give a
reformulation in terms of Schur functions which are determined by
the so-called {\bf Permutation Weights} and there are only a finite
number of permutation weights at each and every order of weight
depths .

Affine signatures are expressed in terms of an index which by
definition is based on a decomposition of horizontal weights in
terms of some {\bf Fundamental Weights}.

\hfill\eject

\vskip 3mm
\noindent {\bf{I.\ INTRODUCTION}}
\vskip 3mm

Although we will consider only $A_r^{(1)}$ algebras in this work, this section
is valid in general for any other affine Lie algebra for which we follow
the notation of Kac {\bf [1]}.

Just as in Weyl character formula {\bf [2]}, in any application of Weyl-Kac
character formula the central idea is again to calculate
$$ A(\La) \equiv \sum_\om \ep(\om) \ e^{\om(\La)}  \eqno(I.1) $$
where $\om$ is an element of affine Weyl group and $\ep(\om)$ is the
corresponding signature. The formal exponentials $e^{\om(\La)}$ are defined
as in the book of Kac {\bf [1]}. This is, however, not our point of
view all along this work. Let $\tr$ and $\La^+$ be, respectively,
affine Weyl vector and an affine dominant weight. Then, we know {\bf [3]}
that (I.1) is equivalent to
$$ A(\tr + \La^+) \equiv \sum_{\La \in W(\tr + \La^+)}  \ep(\La) \ e^{\La}
\eqno(I.2) $$
where $W(\tr + \La^+)$ is the corresponding Weyl orbit. This, hence, forces
us to know the complete weight structure of Weyl orbits of affine dominant
weights. To this end, two main characters are two isotropic elements $\La_0$
and $\dl$ which are defined by
$$ (\La_0,\La_0) = 0 \ \ , \ \ (\dl,\dl) = 0 \ \ , \ \
(\dl,\La_0) = 1  \eqno(I.3)  $$ where the existence of a symmetric
scalar product are always known to be valid. In fact, one gets
affine $\af$ Lie algebra when a finite $A_r$ Lie algebra is
horizontally coupled to system (I.3). Horizontally coupled here
means that
$$ (\dl,\mu) = (\La_0,\mu) = 0   \eqno(I.4) $$
where $\mu$ is an element of $A_r$ Weight Lattice. Any element $\La$ of
$\af$ Weight Lattice can then be expressed in the general form
$$ \La = k \ \dl - M \ \La_0 + \mu \eqno(I.5) $$
where \ k \ is the {\bf level} and \ M \ the {\bf depth} of $\La$.
In (I.5), the level and depth are defined to be non-negative integers.
Note here that the depth is always assumed to be zero for an affine
dominant weight $\La^+$ within an integrable irreducible representation
$R(\La^+)$. As in (I.5), a similar expression
$$ \si(\La^+) = k \ \dl - M_\si \ \La_0 + \mu_\si \eqno(I.6) $$
is valid for any element $\si(\La^+)$ of affine Weyl orbit
$W(\La^+)$ where $\si$ is an affine Weyl reflection, that is an
element of affine Weyl group $W(\af)$. Note also that the level is
always the same for all weights of $R(\La^+)$ and that $\mu_\si$
is again horizontal. A definitive condition here is known to be
$$ (\si(\La^+),\si(\La^+)) = (\La^+,\La^+) \ \ . \eqno(I.7) $$

\noindent (I.7) is, however, insufficient in many cases one of which
is the case we studied here. At this point, the so-called
permutation weights help us to determine all the weights
participating in the same affine Weyl orbit completely. This will be
clarified in section.III in which we give a constructive lemma for
permutation weights. In view of this lemma, it will be seen that for
the same affine Weyl orbit, there are always a finite number of
permutation weights with the same depth and these permutation
weights determine the contributions to Weyl-Kac formula. This, in
fact, reflects our main point of view here in this work.

To be more specific, if one express the right-hand side of Weyl-Kac formula by
$$\sum_M^\infty \ C(M) \ q^M   \ \ , $$
contributions to $C(M_0)$ for any fixed value $M_0$ could come only from
permutation weights having the same value $M_0$. This means that
we can control the contributions to Weyl-Kac character formula at each and
every finite order M individually. It is clear that this is contrary to the
one's expectation because in applications, which are based on Kac-Peterson
form of affine Weyl groups, there is a summation over a part of the whole
root lattice and hence, for instance, two different roots with the same
length could in general contribute in more than one C(M). This will also be
clarified by an example in section.IV.

\vskip 3mm
\noindent {\bf{II.\ CALCULATION OF $A_r$ CHARACTERS}}
\vskip 3mm

It will be useful to give here a brief discussion of how we consider
$A_r$ Lie algebras. This provides us a comparative look on results
we obtained for both affine $\af$ and its finite horizontal algebra
$A_r$. For finite Lie algebras, we refer to the excellent book of
Humphreys {\bf [4]}.

Let, for $i=1,2 \ldots r$, $\al_i$'s and $\la_i$'s be simple roots
and fundamental dominant weights of $A_r$, respectively. Our essential
point of view is again to use {\bf fundamental weights} \ $\mu_I$ \

\hfill\eject
\noindent $(I=1,2 \ldots r+1)$ \ as building blocks. We define the system
$\{ \al_i \}$ of simple roots
$$ \al_i \equiv \mu_i - \mu_{i+1}  \eqno(II.1) $$
\noindent as being in line with the following $A_r$ Dynkin diagram:

\midinsert \epsfxsize=6cm \centerline{\epsfbox{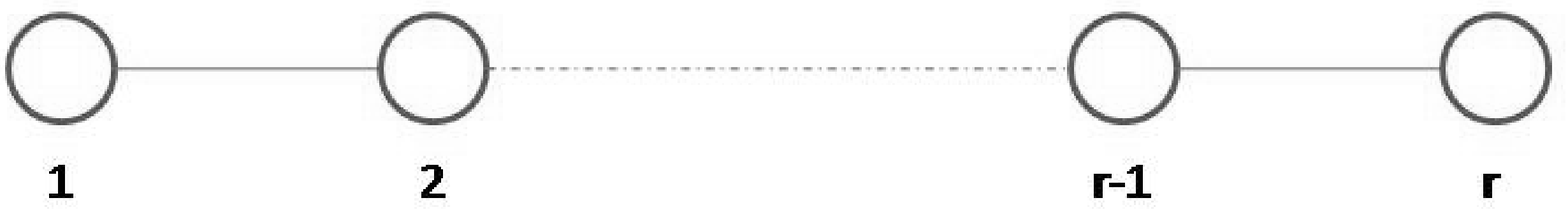}}
\endinsert

\noindent Together with the condition
$$ \sum_{I=0}^{r+1} \ \mu_I \equiv 0 \ \ , \eqno(II.2) $$
(II.1) gives us a complete definition of our fundamental weights. For only
$A_r$ Lie algebras fundamental weights become the weights of (r+1)-dimensional
fundamental representation $R(\la_1)$. This is however not the case for
any other finite Lie algebra.

For $s=1,2, \ldots r$, any dominant weight $\la^+$ then has the form
$$ \la^+ = q_1 \ \mu_1 + q_2 \ \mu_2 + \ldots + q_s \ \mu_s  \eqno(II.3) $$
where
$$ q_1 \geq q_2 \geq \ldots \geq q_s > 0 \ \ .   \eqno(II.4) $$
(II.3) and (II.4), show that there is always a unique partition
$$ \{ \la^+ \} \equiv \{ q_1 , q_2  \ldots  q_s \} \eqno(II.5) $$
which can be defined naturally for any $A_r$ dominant weight $\la^+$.
It is, in fact, nothing but a partition of
$$ h(\la^+) \equiv q_1 + q_2 + \ldots + q_s  \eqno(II.6) $$
which is known to be the {\bf height} of $\la^+$. Then, for any partition
(II.5), one can define a Schur function
$$ S_{\{\la^+ \}}(x_1,x_2,..,x_r) = Det \pmatrix{
S_{q_1-0}&S_{q_1+1}&S_{q_1+2}&\ldots&S_{q_1+s-1}\cr
S_{q_2-1}&S_{q_2+0}&S_{q_2+1}&\ldots&S_{q_2+s-2}\cr
S_{q_3-2}&S_{q_3-1}&S_{q_3+0}&\ldots&S_{q_3+s-3}\cr
\vdots&\vdots&\vdots&\vdots\cr}  \eqno(II.7)  $$
where Det means determinant and $x_i$'s are some indeterminates given below.
For convenience, we keep here the notation
$ S_q \equiv S_{\{q \ \la_1\}}(x_1,x_2, \ldots ,x_r) \ \ .   $

When one considers, on the other hand, (I.1) for $A_r$ Lie algebras,
it is known that
$$ A(\rho+\Lambda^+) = Det \pmatrix{
u_1^{q_1+r}&u_1^{q_2+r-1}&\ldots&u_1^{q_{r+1}}\cr
u_2^{q_1+r}&u_2^{q_2+r-1}&\ldots&u_2^{q_{r+1}}\cr
\vdots&\vdots&\vdots&\vdots&\cr
u_{r+1}^{q_1+r}&u_{r+1}^{q_2+r-1}&\ldots&u_{r+1}^{q_{r+1}} \cr} \eqno(II.8) $$
in the specialization
$$ e^{\mu_I} = u_I \ \ , \ \ I=1,2 \ldots r+1  \eqno(II.9) $$
of formal exponentials.
With the aid of substitution
$$ x_i \equiv {u_1^i + u_2^i + \ldots + u_{r+1}^i \over i}  \eqno(II.10)$$
in (II.8), one can show that
$$ A(\rho+\la^+) = A(\rho) \ S_{\{ \la^+\}}(x_1,x_2 \ldots x_r) \ \ .
\eqno(II.11)$$
This last expression is of central importance in this work and for a
discussion of how one reaches (II.11) we refer to a previous work {\bf [5]}
in which we also give a very detailed study of an $A_5$ example.

\vskip 3mm
\noindent {\bf{III.\ A CALCULATION OF $\af$ CHARACTERS}}
\vskip 3mm

With the inclusion of an extra simple root $\al_0$ defined by
$$ \al_0 \equiv \mu_{r+1} - \mu_1 + \dl  \eqno(III.1) $$
one gets affine $\af$ in view of those given in previous two
sections. The crucial point here is to see, from (I.3), that $\al_0$ is dual
to $\La_0$. This, hence, allows us to obtain an affine system
$$ \{ \al_\nu \ \ , \ \ \nu=0,1,2 \ldots r \} $$
of simple roots for $\af$ whereas, in addition to $\La_0$, a dual system
of affine fundamental dominant weights $\La_\nu$'s can then be defined by
$$ \La_i = \La_0 + \la_i \ \ , \ \ i=1,2 \ldots r \eqno(III.2) $$

\hfill\eject

\noindent with respect to the following extended Dynkin diagram:

\midinsert \epsfxsize=6cm \centerline{\epsfbox{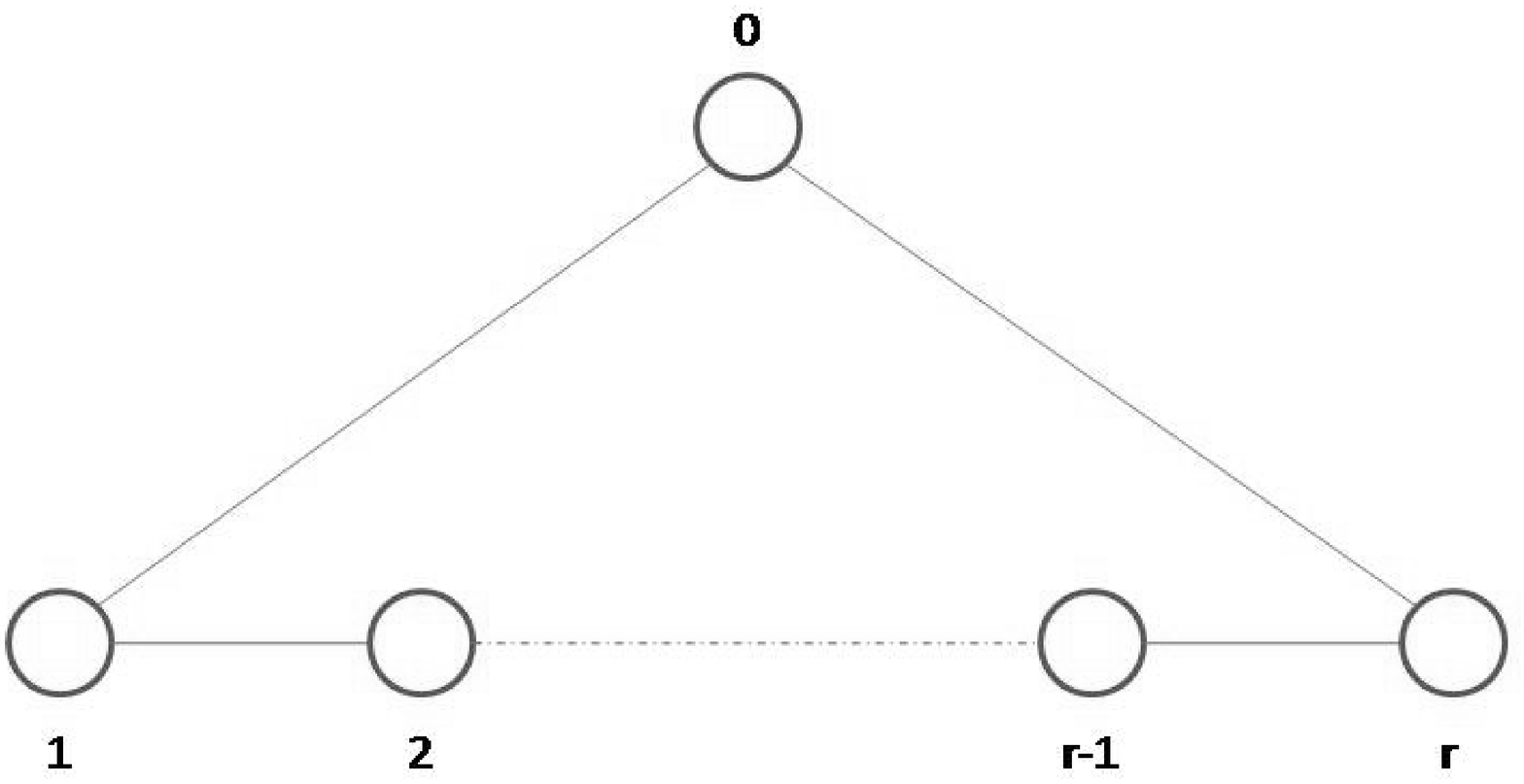}}
\endinsert

\noindent Any other affine dominant weight $\La^+$ of level k then
has the form
$$ \La^+ = k \ \La_0 + \la^+  \eqno(III.3) $$
where $\la^+$ is an $A_r$ dominant weight with $ h(\la^+) \leq k$. It is
known {\bf [1]} that there is an integrable highest weight representation
of $\af$ for each and every affine dominant weight of the form (III.3).
Note here that adjoint representation is not of this type because it has k=0.

As we addressed in section I, (I.1) leaves us together with two main problems
one of which is a complete description of weights participating in the same
affine Weyl orbit. Note here that the problem arises due to the
fact that the condition (I.5), which expresses the invariance of weight lenths
under affine Weyl reflections, becomes sufficient only for affine dominants
with $k=1$ while it remains insufficient in general for a dominant weight
with $k>1$. It is this fact which leads us to state the following
lemma which manifests why we need to define permutation weights.
\medskip
\noindent $\underline{\hbox{DEFINITION:}}$ \ Let
$\La^+ = k \ \La_0 + \la^+$ be $\af$, $\mu^+$ and $\la^+$ $A_r$ dominant
weights. Then, $\si(\La^+)$ is called a {\bf permutation weight}
for $W(\La^+)$ if there is an element $\si \in W(\af)$ such that
$$ \si(\La^+) = k \ \La_0 - M \ \dl + \mu^+ \ \ . \eqno(III.4) $$
\medskip
\medskip

It will be useful to recall here that, as all other elements of the same
Weyl orbit, any permutation weight solves the equation
$$ (\mu^+,\mu^+) - (\la^+,\la^+) = 2 \ k \ M \ \ . \eqno(III.5) $$
Let us also remark here that by saying an $A_r$ dominant weight $\mu^+$
solves (III.5) we mean there is a corresponding permutation weight
as in (III.4). This will be useful in the understanding of the lemma
given below. To obtain a complete list of permutation weights we have now
two steps one of which, the corollary, is for only affine fundamental
dominant weights $\La_\nu$.
\medskip

\noindent $\underline{\hbox{COROLLARY:}}$ \ Let, for an affine dominant
weight $\La^+$, $\PD(\La^+,M)$ be the set of permutation weights with
depths $\leq$ M. Then,

\noindent (1) any solution of (III.5) is in $\PD(\La_\nu,M)$,

\noindent (2) $\PD(\La_\nu,M)$ is finite dimensional for finite M .

The second part of this corollary can be easily seen by noting that
one always has
$$ \mu^+ = \la^+ + \sum_{i=1}^r \ n_i \ \al_i  $$
where $n_i$'s are some non-negative integers. It is clear that only a finite
number of sets $\{ n_1,n_2 \ldots n_r \}$ can solve (III.5) and hence, at each
and every order M of weight depths, there is only a finite number of
contributions to Weyl-Kac formula.

For any other affine dominant weight (III.3), the problem of a complete
determination of the corresponding $\PD(\La^+,M)$  will then be provided
by the following lemma.
\medskip
\noindent $\underline{\hbox{LEMMA:}}$ \ For an affine dominant
$\La^+ = \La_{\nu_1} + \La_{\nu_2} + \ldots + \La_{\nu_n}$, any $\mu^+$
which solves (III.5) is in $\PD(\La^+,M)$ on condition that
$$ \mu^+ \equiv \theta_1 + \theta_2 + \ldots + \theta_n  \eqno(III.6)   $$
where $\theta_1 \in \PD(\La_{\nu_1},M_1) \ ,
\ \theta_2 \in \PD(\La_{\nu_2},M_2) \ , \ \ldots \ ,
\ \theta_n \in \PD(\La_{\nu_n},M_n)$
and
$$ M = M_1 + M_2 + \ldots + M_n \ \ .  $$
providing $M \geq M_j$ for $j=1,2, \ldots n.$
\medskip
\medskip

The second problem mentioned at the beginning of this section is
to determine signatures $\ep(\si)$ for an infinite number of affine Weyl
reflection $\si$. We instead find useful to solve this problem
by defining a new index which gives us the precise values of signatures
$ \ep(\La) $ for any $\La \in W(\rho+\La^+)$.
The fundamental weights $\mu_I$ will play again a major role in the
construction of such an index for signatures.
\medskip
\noindent $\underline{\hbox{DEFINITION:}}$ \ Let
$\ep(s_1,s_2, \ldots ,s_{r+1})$ be completely antisymmetric with the
normalization
$$ \ep(s_1,s_2, \ldots ,s_{r+1}) = +1  \ \ , \ \ s_1 \geq s_2 \geq \ldots
\geq s_{r+1} \ \ . \eqno(III.7) $$
Let $n_I$'s be non-negative integers and $s_I$'s be some integers taking
their values from the finite set $\{0,1,2 \ldots k\}$. Let also define
$$ \ep \ \ \ : \ \ \ \sum_{I=1}^{r+1} \ (s_I + k \ n_I) \ \mu_I
\rightarrow \ep(s_1,s_2, \ldots ,s_{r+1}) \ \ . \eqno(III.8) $$
\medskip
\medskip
Note here that $\ep$ is a mapping from $A_r$ weight lattice to
$Z_2 \equiv \{+1,-1\}$. In view of the fact that $h(\mu^+) \leq k$, it is
obvious that any permutation weight $\mu^+$ has a decomposition of the form
$$ \mu^+ = \sum_{I=1}^{r+1} \ (s_I + k \ n_I) \ \mu_I  \ \ . \eqno(III.9) $$
Its signature $\ep(\mu^+)$ then will be equal to the value of the mapping
(III.8) on the point $\mu^+$ of $A_r$ Weight lattice.

All these, when they are considered together with those in ref.{\bf
[5]} concerning $A_r$ Lie algebras, give us the possibility to
express (I.1) in a form which is our ultimate result here in this
work:
$$  A(\La^{++},M) \ \ \
\vtop{\hbox{$\arx$} \hbox{${\scriptscriptstyle M \to \infty}$}}
\ \ \ A(\La^{++}) \eqno(III.10) $$
where
$$ A(\La^{++},M) \equiv A(\rho) \ \sum_{\mu^+ \in \PD(\La^{++},M)} \
S_{\{\mu^+-\rho\}}(x_1,x_2 \ldots x_{r+1})  \eqno(III.11) $$
It is therefore clear that, for any chosen value $M_0$ of M, (III.11)
determines the contributions to Weyl-Kac formula up to order $M_0+1$.
An example will be given in the next section.

\vskip 3mm
\noindent {\bf{IV. FROM TETA FUNCTIONS TO SCHUR FUNCTIONS}}
\vskip 3mm

Our main exposition in this work is to express affine characters in terms of
Schur functions just as in case of finite Lie algebras. This, on the other hand,
could seem to be contrary to the one's expectation because in the conventional
scheme affine characters are expressed in terms of teta functions. In the light
of the work of Kac and Peterson {\bf [6]}, this is natural because affine
Weyl groups include some kind of translations. Beyond that, another
characteristic difference between the two approaches comes out in actual
computations of affine characters. To this end, a brief discussion of the
interplay between Weyl-Kac formula and teta functions will be useful. The
discussion below considered only for $A_r^{(1)}$ Lie algebras though it could
be easily generalized for any affine Lie algebra.

In its classical treatment, the Jacobi's teta function is known to be
defined by
$$ \theta_3(\tau,z) \equiv \sum_{n=-\infty}^{\infty} e^{\pi i \tau n^2} \
e^{2 \pi i z n} \eqno(IV.1)  $$
which allows us to write
$$ \theta_3({\scriptstyle \tau},{s \over {\scriptstyle r+1}}) \equiv 1 \ + \
\sum_{n=1}^\infty \ (u(s n) + u(-s n)) \ q^{{\scriptstyle {1 \over 2}} n^2} $$
where $q \equiv e^{2 \pi i {\scriptstyle \tau}}$ and formal exponentials
$$ u(s) \equiv e({2 \pi i \over {\scriptstyle r+1}} s) \ \ , \ \
s = \mp 1, \mp 2, \ldots $$
can be defined in such a way that the limits $ u(s) \to 0 $ is valid formally
for all values of integers s, positive or negative. One easily shows then
that the lattice function $\Theta(A_r)$ of root lattice $\Gamma(A_r)$
is expressed, for $i=1,2, \ldots r$, by
$$ \hbox{$ \theta_3({\scriptstyle \tau},{i \over
{\scriptstyle r+1}})^{^{\scriptstyle r+1}}$ \ \
\vtop{\hbox{$\ary$} \hbox{${\scriptstyle u(s) \to 0}$}}
\ \ $\Theta(A_r)$ } \eqno(IV.2)  $$

Let us remark here that the more usual formulation of $\Theta(A_r)$ can be
found in the book of Conway and Sloane {\bf [7]}. Now, if one recalls
$$ \Gamma(A_r) = \bigcup_{n=0}^\infty \Gamma_n $$
where
$$ \Gamma_n \equiv \{ \al \ \vert \ (\al,\al)=2 n \} \ \ ,  \eqno(IV.3) $$
the definition
$$ \Theta(A_r) \equiv \sum_{n=0}^\infty \ q^n \ dim\Gamma_n \eqno(IV.4) $$
is valid. One has, for instance,
$$ \Theta(A_4) = 1 + 20 \ q + 30 \ q^2 + 60 \ q^3 + 60 \ q^4 + 120 \ q^5 +
40 \ q^6 + 180 \ q^7 + 150 \ q^8 + \ldots \eqno(IV.5) $$

For $\af$ Lie algebras, on the other hand, a natural extension
$$ \theta_3(\tau, z) \ \arx \ \theta_\La(\tau,z_1,z_2, \ldots ,z_r) \equiv
\theta_\La \eqno(IV.6) $$
is valid for any affine weight $\La$ and some coordinates $z_i$. Let us define
$$ \theta_\La \equiv \sum_{\al \in \Gamma(A_r)} \
q^{{1 \over 2 k} (\La+k \al,\La+k \al)} \
\prod_{i=1}^r u_i^{(\La+k \al,\al_i)} \eqno(IV.7) $$
where $u_i \equiv e^{2 \pi i z_i}$ and
$$ k \equiv (\La,\delta) \ \ .  \eqno(IV.8) $$
It is clear that (IV.7) is a kind of teta function. Its convenience,
however, more clearly understood when one recalls that there is an appropriate
spcecialization of Weyl-Kac character formula so that affine characters
$Ch(\La^+)$ are expressed by the aid of
$$ B(\La^+) \equiv \sum_{\om \in W(A_r)} \ep(\om) \ \theta_{\om(\La^++\tr)}
\ \ . \eqno(IV.9) $$
What is remarkable in (IV.9) is that the sum is over
the finite Weyl group $W(A_r)$. This is a result due to Kac-Peterson.
(IV.9) can be cast into the form
$$ B(\La^+) \equiv \sum_{\al \in \Gamma_n} B(n,\La^+) \eqno(IV.10) $$
where $B(n,\La^+)$ can be extracted from definition (IV.7). An equivalent
form can be given, however, as in the following:
$$ B(n,\La^+) = q^{a(A_r,\tk)} \ B(0,0^{\scriptscriptstyle+}) \ \
\ct_{n,\La^+}(q,u_1,u_2, \ldots ,u_r) \eqno(IV.11) $$
where $\ct_{n,\La^+}(q,u_1,u_2, \ldots ,u_r)$ is a
polinomial with integer powers of indeterminate q and $a(A_r,\tk)$
is the so-called anomaly which is known to be defined by
$$ a(A_r,\tk) \equiv {1 \over 2 \ \tk} \
\big( (\La^+,\La^++2 \ \tr) - {1 \over 12} \ \tk \ dimA_r \big) \eqno(IV.12) $$
where $ \tk \equiv (\La^++\tr,\delta)$ and $dimA_r=r(r+2)$. It is therefore
seen that when one attempts to obtain affine characters the main problem
is to calculate polinomials $\ct_{n,\La^+}$ \ . Weyl-Kac formula then gives
us {\bf normalized characters}
$$ \chi_{\scriptstyle \La^+}(\tau,z_1,z_2, \ldots ,z_r) =
{{1 + \ct_{1,\La^+} + \ct_{2,\La^+} + \ldots} \over
{1 + \ct_{1,0^{\scriptscriptstyle+}} + \ct_{2,0^{\scriptscriptstyle+}} +
\ldots} } \ \ . \eqno(IV.13) $$
(IV.13) never permits to a complete factorization which means in effect that
it consists of an infinite sum of powers of q. Although in a few simple cases
first few elements lead us to compact expressions {\bf [9]}, normalized
characters can be given only up to some order $N < \infty $, in general.
It is clear that this brings out an unexpected problem in the theory because,
as we stressed above, the polinomials $\ct_{n,\La^+}$ contain in general more
than one terms with different powers of q. Let us first proceed with an
example to illustrate the problem though we will give below how we consider
the problem with the introduction of permutation weights.

One of the specialties of Affine Lie Algebras is the occurrence of basic
representation $R(\La_0)$ having the normalized character {\bf [8]}
$$ \chi_{{\scriptscriptstyle \La_0}}(\tau) =
{ \Theta(A_r) \over \Phi(\tau)^r}  \eqno(IV.14)    $$
where
$$ \Phi(\tau) \equiv \prod_{n=1}^\infty \ (1-q^n) \ \ .  $$
For convenience, we adopt the notation
$$ \chi_{{\scriptscriptstyle \La_0}}(\tau) =
\chi_{{\scriptscriptstyle \La_0}}(\tau,z_1=1,z_2=1, \ldots z_r=1) $$
in the specialization used in (IV.14).

To be more abstract, let us proceed with $A_4$ for which following
contributions to (IV.13) are obtained:
$$ \eqalign{
\ct_{1,0^{\scriptscriptstyle+}} =
&-24 \ q + 252 \ q^2 - 1472 \ q^3 + 3654 \ q^4 - \cr
&19096 \ q^6 + 40128 \ q^7 - 34398 \ q^8 + 10976 \ q^9 \ \ ,  \cr
\ct_{2,0^{\scriptscriptstyle+}} = \
&1176 \ q^4 - 6048 \ q^5 + 2352 \ q^6 + 44352 \ q^7 - 83997 \ q^8 - \cr
&78848 \ q^9 + 360756 \ q^{10} - 157248 \ q^{11} - 530222 \ q^{12} + \cr
&598752 \ q^{13} + 123552 \ q^{14} - 448448 \ q^{15} + 173901 \ q^{16} \ \ , \cr
\ct_{1,\La_0} =
&-200 \ q^2 + 2100 \ q^3 - 9625 \ q^4 + 19096 \ q^5 - \cr
&70200 \ q^7 + 128625 \ q^8 - 98000 \ q^9 + 28224 \ q^{10}  \ \ , \cr
\ct_{2,\La_0} = \
&9000 \ q^6 - 46200 \ q^7 + 23100 \ q^8 + 286000 \ q^9 - 530222 \ q^{10} - \cr
&409500 \ q^{11} + 1851850 \ q^{12} - 754600 \ q^{13} - 2281500 \ q^{14} + \cr
&2354352 \ q^{15} + 477750 \ q^{16} - 1528800 \ q^{17} + 548800 \ q^{18} \ \ . }
\eqno(IV.15)   $$
One sees now that
$$ {1 + \ct_{1,\La_0} \over 1 + \ct_{1,0^{\scriptscriptstyle+}}} =
1 + 24 \ q + 124 \ q^2 + \ldots \eqno(IV.16)  $$
and
$$ \eqalign {
{1 + \ct_{1,\La_0} + \ct_{2,\La_0} \over 1 + \ct_{1,0^{\scriptscriptstyle+}} +
\ct_{2,0^{\scriptscriptstyle+}} } =
&1 + 24 \ q + 124 \ q^2 +  \cr
&500 \ q^3 + 1625 \ q^4 + 4752 \ q^5 + 12524 \ q^6 +
31000 \ q^7 + \ldots   }  \eqno(IV.17)  $$
reproduces (IV.14) up to third and eigtheeth degree correctly. Both are
insufficient to reproduce a correct $q^8$ term though all the polinomials
in (IV.15) have non-zero $q^8$ terms. We need, in fact, third order
contributions to calculate $\chi_{\La_0}$ beyond $q^7$.

We will now show that we only need permutation weights $\PD(\La^+,2)$ and
$\PD(\La^+,7)$ in order to calculate (IV.16) and (IV.17) respectively.
In view of the lemma given above one indeed has
$$ \eqalign { \PD(\tr,8) = \{
&(0,0,0,0)_0 \ , \ (1,0,0,1)_1 \ , \ (2,0,1,0)_2 \ , \
(0,1,0,2)_2  \ , \ (3,1,0,0)_3 \ , \cr
&(1,1,1,1)_3 \ , \ (0,0,1,3)_3  \ , \ (5,0,0,0)_4 \ , \
(2,2,0,1)_4 \ , \ (1,0,2,2)_4 \ , \cr
&(0,2,2,0)_4 \ , \ (0,0,0,5)_4  \ , \ (1,3,1,0)_5 \ , \
(0,1,3,1)_5 \ , \ (1,0,0,6)_6 \ , \cr
&(0,5,0,0)_6 \ , \ (2,0,2,3)_6 \ , \ (3,2,0,2)_6 \ , \
(6,0,0,1)_6 \ , \ (0,0,5,0)_6 \ , \cr
&(2,0,1,5)_7 \ , \ (3,1,1,3)_7 \ , \ (1,1,3,2)_7 \ , \
(5,1,0,2)_7 \ , \ (2,3,1,1)_7 \ , \cr
&(3,1,0,5)_8 \ , \ (1,5,0,1)_8 \ , \ (0,1,0,7)_8 \ , \
(5,0,1,3)_8 \ , \ (2,2,2,2)_8 \ , \cr
&(1,0,5,1)_8 \ , \ (7,0,1,0)_8  \}  }  $$

$$ \eqalign { \PD(\tr + \La_0,8) = \{
&(0,0,0,0)_0 \ , \ (2,0,0,2)_2 \ , \ (3,0,1,1)_3 \ , \
(1,1,0,3)_3 \ , \ (4,1,0,1)_4 \ , \cr
&(2,1,1,2)_4 \ , \ (1,0,1,4)_4 \ , \ (6,0,0,1)_5 \ , \
(3,2,0,2)_5 \ , \ (2,0,2,3)_5 \ , \cr
&(1,0,0,6)_5 \ , \ (0,3,3,0)_6 \ , \ (1,4,2,0)_7 \ , \
(0,2,4,1)_7 \ , \ (2,0,0,7)_7 \ , \cr
&(3,0,2,4)_7 \ , \ (4,2,0,3)_7 \ , \ (7,0,0,2)_7 \ , \
(0,6,1,0)_8 \ , \ (0,1,6,0)_8 \ , \cr
&(3,0,1,6)_8 \ , \ (4,1,1,4)_8 \ , \ (6,1,0,3)_8 \} } $$
where the notation
$$ k \ \La_0 - M \ \delta + \rho + \sum_{i=1}^4 \ n_i \mu_i \equiv
(n_1,n_2,n_3,n_4)_{\scriptscriptstyle M} $$
is used for convenience. Since  by definition
$$ \PD(\La^+,n) \supset \PD(\La^+,m) \ \ \ , \ \ \ n>m \ \ , $$
this means in effect that
$$ \PD(\La^+,8) \supset \PD(\La^+,7) \supset \PD(\La^+,2)  \ \ .  $$
Now if one conveniently defines
$$ \chi_{\scriptscriptstyle \La_0}(\tau,M) \equiv
{ \sum_{\mu^+ \in \PD(\tr+\La_0,M)} \ S_{\{\mu^+\}}(x_1,x_2 \ldots x_5)
\over \sum_{\mu^+ \in \PD(\tr,M)} \ S_{\{\mu^+\}}(x_1,x_2 \ldots x_5) } \ \ .
\eqno(IV.18)  $$
then one sees that $ \chi_{\scriptscriptstyle \La_0}(\tau,2)$ and
$ \chi_{\scriptscriptstyle \La_0}(\tau,7)$ give respectively the terms
explicitly shown in (IV.16) and (IV.17) while
$ \chi_{\scriptscriptstyle \La_0}(\tau,8)$ reproduces (IV.14) up to
nineth order.

\vskip 3mm
\noindent {\bf{IV. CONCLUSION}}
\vskip 3mm

We extend the concept of permutation weights, which we have introduced
for finite Lie algebras previously, to affine Lie algebras. This brings us
out the formula (III.11) which permits a reformulation of Weyl-Kac character
formula. In comparison of (III.11) with (IV.9), our emphasis is that the sum
over Weyl groups is removed in the former whereas explicit calculation of
permutation weights is a simple task even for the huge algebra of $E_8$
Lie group. The hope however is that the permutation weights would have
a similar meaning also for hyperbolic Lie algebras.

\vskip3mm
\noindent{\bf {REFERENCES}}
\vskip3mm

\item{[1]} V.G.Kac, Infinite Dimensional Lie Algebras, N.Y.,
Cambridge Univ. Press (1990)

\item{[2]} H.Weyl, The Classical Groups, N.J. Princeton Univ. Press (1946)

\item{[3]} H.R.Karadayi and M.Gungormez, J.Phys.A:Math.Gen. 32 (1999) 1701-1707

\item{[4]} J.E.Humphreys, Introduction to Lie Algebras and Representation Theory, \hfill\break
N.Y., Springer-Verlag (1972)

\item{[5]} H.R.Karadayi, $A_N$ Multiplicity Rules and Schur Functions,
math-ph/9805009

\item{[6]} V.G.Kac and D.H.Peterson, Advances in Math,. 53 (1984), 125-264

\item{[7]} J.H.Conway and N.J.A.Sloane, Sphere Packings,Lattices and Groups, \hfill\break
chapter 4, N.Y., Springer-Verlag (1991)

\item{[8]} A.J.Feingold and J.Lepowsky, Advances in Math,. 29 (1978), 271-309 , \hfill\break
V.G.Kac, Advances in Math,. 30 (1978), 85-136

\item{[9]} V.G.Kac and D.H.Peterson, Bull.Amer.Math.Soc. 3 (1980) 1057-1061

\end